\documentclass[prl, onecolumn, floatfix, amsmath]{revtex4}
\usepackage{float}

\usepackage{graphicx}
\usepackage{epstopdf}
\expandafter\let\csname equation*\endcsname\relax
\expandafter\let\csname endequation*\endcsname\relax
\usepackage{amsmath}
\usepackage{amssymb}
\usepackage{amsthm}
\usepackage[section]{placeins}
\usepackage[UKenglish]{datetime}
\usepackage[english]{babel}
\usepackage{url}

\usepackage[pdftex,colorlinks]{hyperref}
\usepackage[all]{hypcap}
\hypersetup{colorlinks=true,citecolor=black, linkcolor=black, urlcolor=blue, unicode=true}

\begin{document}

\title{Superpixel-based spatial amplitude and phase modulation using a digital micromirror device} 

\author{Sebastianus A. Goorden}\email{s.a.goorden@utwente.nl}
\affiliation{Complex Photonic Systems (COPS), MESA+ Institute for Nanotechnology, University of Twente, P.O. Box 217, 7500 AE Enschede, The Netherlands}

\author{Jacopo Bertolotti}
\affiliation{Complex Photonic Systems (COPS), MESA+ Institute for Nanotechnology, University of Twente, P.O. Box 217, 7500 AE Enschede, The Netherlands}
\affiliation{Physics and Astronomy Department, University of Exeter, Stocker Road, Exeter EX4 4QL, United Kingdom}

\author{Allard P. Mosk}
\affiliation{Complex Photonic Systems (COPS), MESA+ Institute for Nanotechnology, University of Twente, P.O. Box 217, 7500 AE Enschede, The Netherlands}



\begin{abstract}
We present a superpixel method for full spatial phase and amplitude control of a light beam using a digital micromirror device (DMD) combined with a spatial filter. We combine square regions of nearby micromirrors into superpixels by low pass filtering in a Fourier plane of the DMD. At each superpixel we are able to independently modulate the phase and the amplitude of light, while retaining a high resolution and the very high speed of a DMD. The method achieves a measured fidelity $F=0.98$ for a target field with fully independent phase and amplitude at a resolution of $8\times 8$ pixels per diffraction limited spot. For the LG$_{10}$ orbital angular momentum mode the calculated fidelity is $F=0.99993$, using $768\times 768$ DMD pixels. The superpixel method reduces the errors when compared to the state of the art Lee holography method for these test fields by $50\%$ and $18\%$, with a comparable light efficiency of around $5\%$. Our control software is publicly available. 
\end{abstract}


\maketitle

\section{Introduction}
Full control over light allows many exciting applications. By tailoring light fields we can now use optics to obtain a great level of control over particles \cite{Denz2013_lasphotrev}. Shaping light waves greatly improves our ability to see the world around us through optical microscopy \cite{RitschMarte2011_review,Bhaduri2014,mertz_2012,Fleischer2012_natphot} and allows exciting technologies in the field of optical communication, crucial to support the quantity and security of the rapidly expanding amount of information that is sent around the world \cite{Gibson_optexpress}. Wavefront shaping allows compensation for and exploitation of scattering due to spatial inhomogenieties in the refractive index of a material \cite{mosk_2012_naturephoton}. In this way it is possible to image through \cite{Vellekoop2007_optlett,Putten2011_PRL} and inside \cite{Vellekoop2008_optexpr,psaltis2010_optexpress,Wang2011_natphot,yang2012_naturecomm,cui_2012_naturephoton} opaque materials, which is of great importance in biomedical imaging. Light propagating through an opaque material can be controlled in time by spatially shaping the incident wavefront \cite{Aulbach2011_prl,Katz2011_nphot,mccabe2011_naturecomm} with applications such as pulse compression. Wavefront shaping also allows the use of multiple-scattering media as a tunable wave plate \cite{park2012_optexp, silberberg2012_optlett}, spectral filter \cite{park2012_optlett, small2012_optlett} or tunable beamsplitter \cite{huisman_optexp_2014}. The digital micromirror device (DMD) \cite{Dudley2003} is an excellent candidate for controlling light fields, as it has a very high number of spatial degrees of freedom, a very high framerate, it operates in a broad wavelength range and it is relatively cheap. Each pixel of a DMD is a mirror which can be in one of two positions, corresponding to the `on' and `off' states of the pixels. Wavefronts can be controlled using binary amplitude modulation \cite{Akbulut2011_optexpr}, but less efficiently than using phase modulation \cite{Vellekoop2007_optlett}. Shaping complex fields with binary masks is of continuous interest \cite{Brown1969,Kreis2001,Ulusoy2011b}, adding to the momentum of the rapidly growing field of computer-generated holography. For reviews see \cite{Lee1978, Tricoles1987, Nehmetallah2012}. The most common technique to obtain phase modulation with a DMD is Lee holography \cite{Lee1974} and has been shown to allow for efficient and fast wavefront shaping \cite{piestun2012highspeed_optexp}. Lee holography in its more general form \cite{Lee1974} allows full field control \cite{Mirhosseini2013}. Lee holography with pixel dithering has been demonstrated and the obtained errors are at the 5$\%$ level for low resolution fields \cite{Lerner2012}. A method has been proposed, but not yet demonstrated, that is based on a complex high spatial resolution Fourier mask and is from an information theoretic point of view optimal \cite{Ulusoy2011}, but requires involved optics and is not robust to misalignment. We propose and demonstrate a superpixel-based \cite{Putten2008} phase and amplitude modulation method, which is highly robust and easy to use while offering full spatial control over the phase and amplitude of a light field. The method is applied, through calculations as well as measurements, to two target fields of high practical relevance: the LG$_{10}$ orbital angular momentum mode and a high resolution field with fully independent amplitude and phase. The modulation accuracy of the method is quantified by calculating the fidelity $F=\left| E_{\mathrm{target}}^* E_{\mathrm{obtained}}\right|^2$ and the error $\delta = 1-F$, where $E_{\mathrm{target}}$ is a target field and $E_{\mathrm{obtained}}$ is the field that is obtained using the DMD. The fidelity of the superpixel method is found to be very high in theory as well as in experiments.
 
 \begin{figure}[t]
 \capstart
 \centerline{\includegraphics[width=14cm]{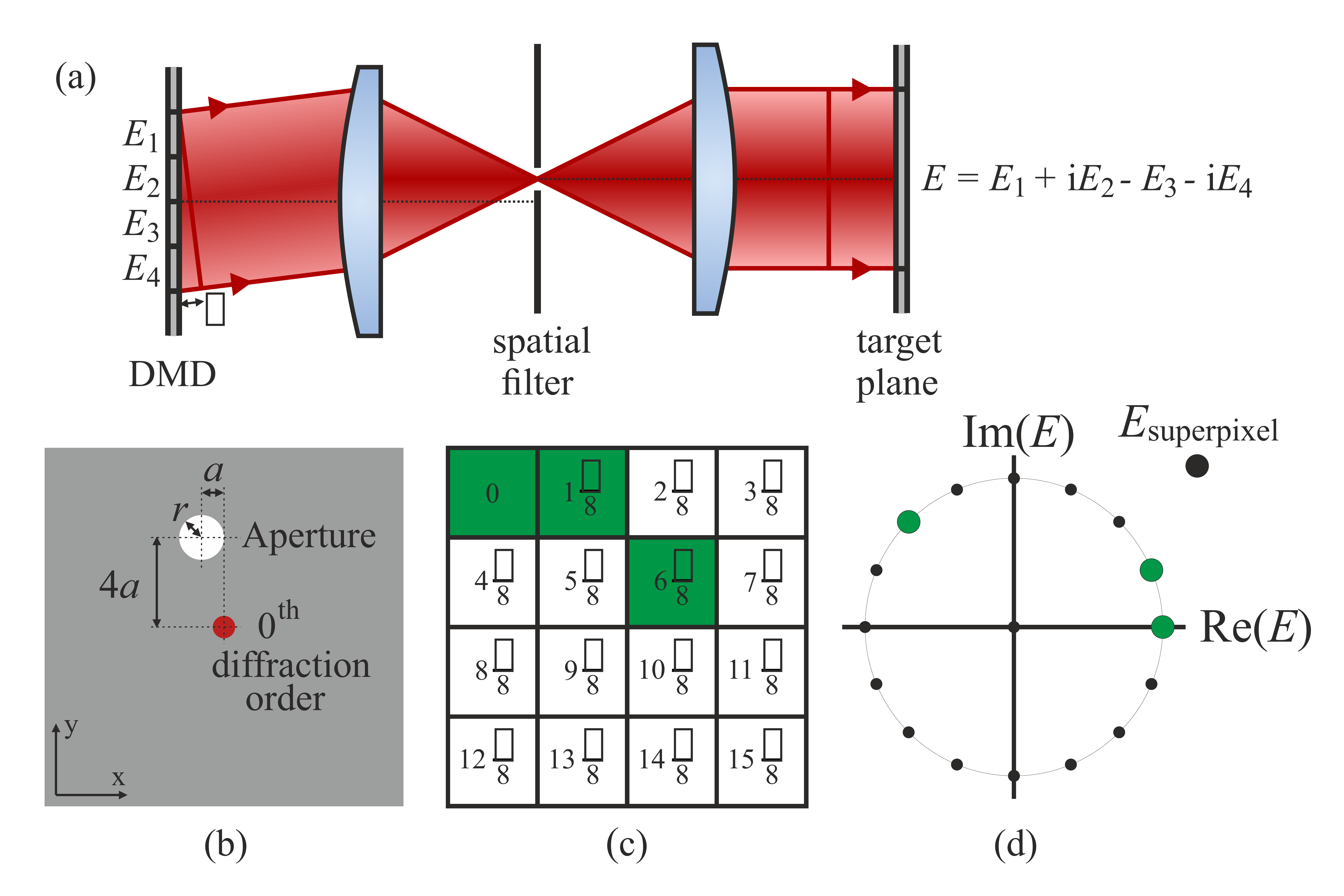}}
 \caption{(a) In the DMD plane the light field $E(\mathbf{x})\in \{0,1\}$, corresponding to the off and on states of the micromirrors. The DMD is imaged onto the target plane in which we maximize the level of control over the light field. The DMD pixel images in the target plane have different phase prefactors, because the lenses are placed off-axis with respect to each other. A low pass filter blurs the images of pixels and averages over groups of neighboring pixels. (b,c,d) The aperture is positioned such that the phase responses of the 16 DMD pixels within a $4\times 4$ superpixel are uniformly distributed between 0 and $2\pi$. Example: if we turn on the three pixels indicated by green squares in (c), then the response $E_{\mathrm{superpixel}}$ in the target plane is the sum of the three pixel responses in (d).}
 \label{fig1}
 \end{figure}

\section{Setup}
Our setup is designed to obtain full spatial control over the phase and amplitude of light in one specific plane, which we call the target plane. The field behind the target plane follows from usual beam propagation methods. Our Vialux V4100 DMD with a resolution of $1024\times 768$ pixels and pixel pitch of $13.68\mu$m is imaged onto the target plane using two lenses in a 4f-configuration, as illustrated in Fig. \ref{fig1}(a). The lenses are placed slightly off-axis with respect to each other, resulting in an extra phase factor in the target plane. This means that the phase of the target plane response of a DMD pixel depends on the position of the pixel on the DMD. The DMD is divided into superpixels: square groups of $n\times n$ micromirrors. The lenses are placed in such a way that the phase prefactors of the micromirrors within each superpixel are distributed uniformly between 0 and $2\pi$. A spatial filter in the form of a circular aperture is placed in the Fourier plane in between the lenses. The spatial filter blocks the high spatial frequencies so that individual DMD pixels cannot be resolved. The images of the pixels in the target plane are blurred and have a large spatial overlap \cite{Putten2008}. Therefore, the target plane response of a superpixel is the sum of the individual pixel responses.

\begin{figure}[t]
\centerline{\includegraphics[width=14cm]{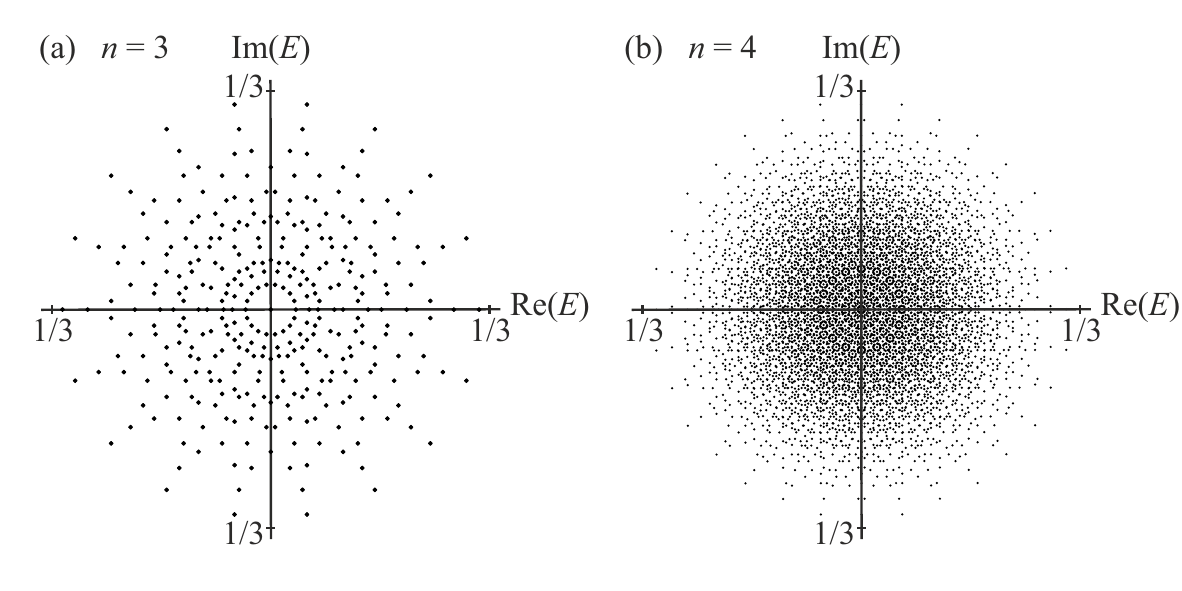}}
\caption{(a) Complex target fields that can be constructed using a single superpixel of size $3\times 3$. 343 different fields can be constructed. (b) Complex target fields that can be constructed using a single superpixel of size $4\times 4$. 6561 different fields can be constructed. Fields are normalized to the incident field. The symbol size is larger for $n=3$ to increase visibility.}
\label{fig2}
\end{figure}

For superpixels of size $n \times n$ the position of the spatial filter with respect to the $0^{\mathrm{th}}$ diffraction order is chosen $(x,y) = (-a,n \ a)$, where $a = \frac{-\lambda f}{n^2 d}$, $\lambda$ is the wavelength of the light, $f$ is the focal length of the first lens and $d$ is the distance between neighbouring micromirrors. This position is chosen such that the target plane responses of neighbouring pixels inside the superpixel are $\frac{2\pi}{n^2}$ out of phase in the $x$-direction and $\frac{2\pi}{n}$ out of phase in the $y$-direction. The target plane responses of the $n^2$ pixels that make up a superpixel are then uniformly distributed over a circle in the complex plane. For superpixels of size $n=4$ this is illustrated in Figs. \ref{fig1}(b)--\ref{fig1}(d). Using our DMD, a HeNe laser with a wavelength of $\lambda = 633$ nm and a first lens with a focal length $f_{1} = 300$ mm, the aperture is positioned at $(x,y) = (-0.87,3.47)$ mm. Therefore, the target plane responses of neighbouring pixels in the $x$ and $y$ direction are $\frac{\pi}{8}$ and $\frac{\pi}{2}$ out of phase, as illustrated in Fig. \ref{fig1}(c). The phase responses in the target plane of the 16 DMD pixels are then distributed uniformly between 0 and $2\pi$, as shown in Fig. \ref{fig1}(d), indicating that we have achieved control over the phase of light. 
 
 As an example, we set a superpixel such that the pixels with phase responses $0$, $\frac{\pi}{8}$ and $\frac{6 \pi}{8}$ are turned on, indicated in Fig. \ref{fig1}(c). All other pixels are turned off. The resulting field $E_{\mathrm{superpixel}}$ in the target plane will be approximately equal to the sum of the three dots in Fig. \ref{fig1}(d). By turning on different combinations of pixels in a superpixel we can create different target fields in the target plane. For all possible combinations of pixels we plot the corresponding target fields in Fig. \ref{fig2}. For superpixels of size $n=3$ we see in Fig. \ref{fig2}(a) that we can construct a total number of 343 different fields, quite uniformly distributed over a disk in the complex plane. For superpixels of size $n=4$ we see in Fig. \ref{fig2}(b) that the number of fields we can construct increases dramatically to 6561, allowing us to create any field within a disk up to a very small discretisation error.

The resolution, or spatial bandwidth, of the superpixel method is given by $\Delta k = \frac{2 \pi r}{\lambda f_{2}} \ \ \textrm{rad}\cdot \textrm{m}^{-1}$, where $r$ is the radius of the aperture and $f_{2}$ is the focal length of the second lens. The target plane is an image plane of the DMD and therefore it is natural to express the resolution in units of DMD pixels: $\Delta k' = \frac{2 \pi d r}{\lambda f_{1}} \ \ \textrm{rad}\cdot \textrm{pixel}^{-1}$. We typically choose $r$ such that our system bandwidth matches the bandwidth of the target field, with an upper limit such that the highest allowed spatial frequency is not higher than $\frac{\pi}{2n}\ \ \textrm{rad}\cdot \textrm{pixel}^{-1}$. This upper limit ensures that the images of DMD pixels are blurred and average out to the desired superpixel field value \cite{Putten2008}. In our system with superpixel size $n=4$ the maximum aperture size is $r=0.9$ mm and the corresponding feature size in the target plane is approximately $2 \times 2$ superpixels.

Alignment of the spatial filter is done in two steps. First, we write a pattern to the DMD which corresponds to a plane wave in the target plane. The spatial filter is placed around the first diffraction order of this grating. Second, we fine-tune the position and size of the spatial filter by writing horizontal and vertical gratings to the DMD that correspond to the desired spatial band limit of the system. We align the spatial filter such that the two diffraction orders of each grating exactly pass through at the edge of the filter. The DMD patterns for these alignment gratings, as well as for any other target field, are calculated using our superpixel control software \cite{superpixel_software}. Using this method accurate alignment of the spatial filter is easily achieved. The effect of misalignment of the filter on the resulting light field depends on the spatial frequency distribution of the target field, but is typically small: e.g. $10\%$ relative displacement of the spatial filter results in less than $0.5\%$ loss of modulation fidelity for test field 2. 

\section{Efficiency, bandwidth and implementation}
The efficiency of the superpixel method is equal to the maximum intensity a superpixel can create. The maximum intensity is obtained by turning on exactly half of the pixels in a superpixel, e.g. the upper 8 pixels in Fig. \ref{fig1}(c). This coincides with the maximum amplitude in Fig. \ref{fig2}(b). The calculated efficiency of the modulation method is then $10.3\%$ of the incident intensity. The measured 0 order diffraction efficiency of the DMD itself is $60\%$ for our DMD and our measured modulation efficiency is $7\%$. The total measured efficiency of our implementation of the superpixel method is $4\%$. This is similar to the efficiency of Lee holography, since both methods are based on filtering out the first order of an intensity diffraction grating. 

\begin{figure}[t]
\center{\includegraphics[width=14cm]{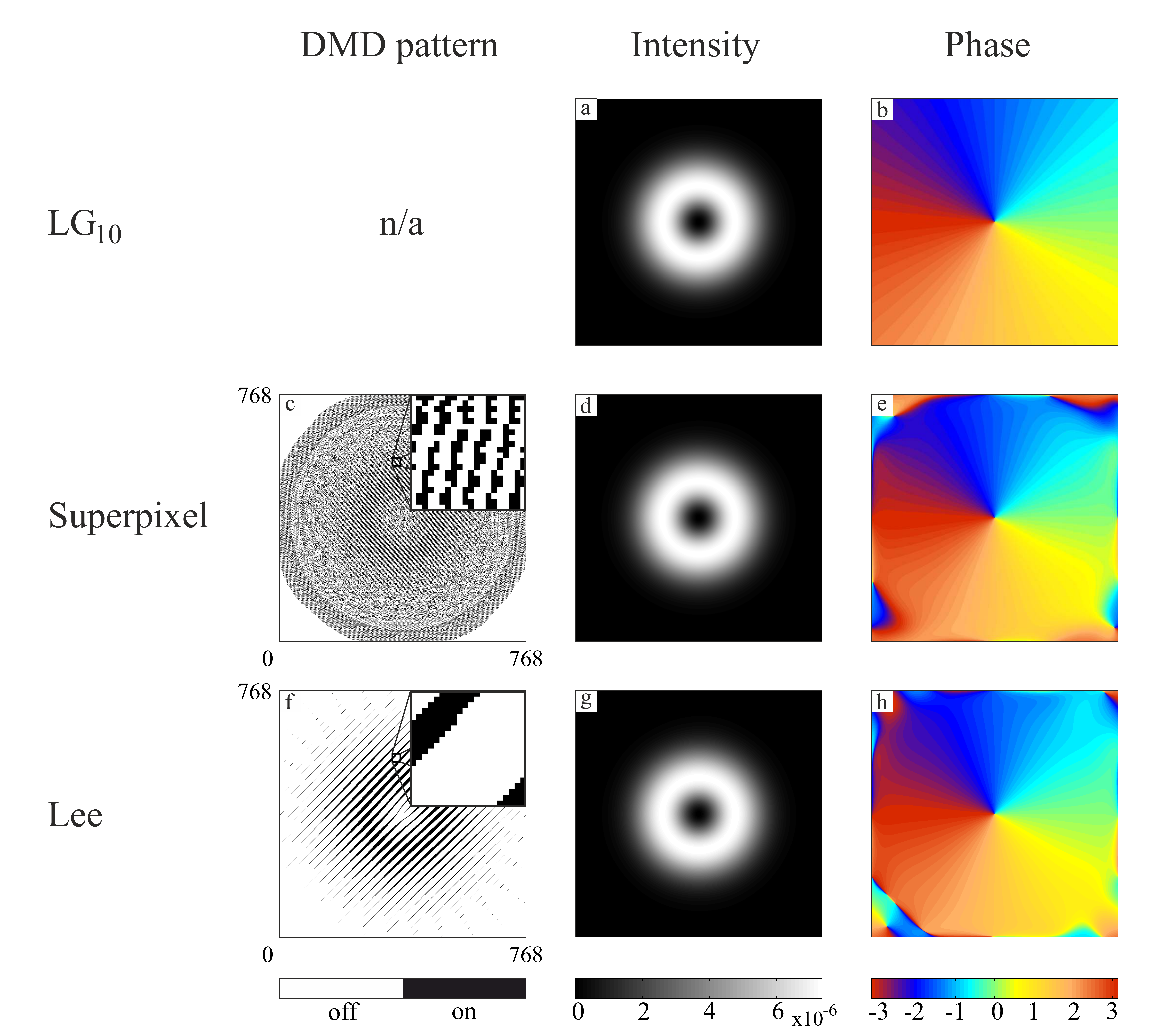}}
\caption{(a,b) Intensity and phase of the target LG$_{10}$ mode. (c) DMD pattern for the LG$_{10}$ mode when using the superpixel method. Inset: zoom-in on $20\times 20$ DMD pixels. (d,e) Calculated intensity and phase using the superpixel method; $\delta_{\mathrm{superpixel}} = 7\cdot 10^{-5}$. (f) DMD pattern for the LG$_{10}$ mode when using Lee holography with $k_{x}=k_{y}=\frac{2 \pi}{30}$ pixel$^{-1}$. Inset: zoom-in on $20\times 20$ DMD pixels. (g,h) Calculated intensity and phase using Lee holography; $\delta_{\mathrm{Lee}} = 9\cdot 10^{-5}$. Intensities are normalized to total intensity.}
\label{fig3}
\end{figure}

Wavelength dispersion of the amplitude mask formed by the DMD limits the frequency bandwidth in which the superpixel method works. The superpixel-based phase and amplitude modulation method can be set up for any wavelength $\lambda$ at which the DMD functions. The position and size of the spatial filter depend on $\lambda$. Illuminating the DMD with light of a different wavelength, e.g. $\lambda + \Delta \lambda$, decreases the modulation fidelity. The error induced in the target plane is, to first order, a phase gradient added to the target field. The period of the phase gradient is equal to $\left| n \lambda / \Delta \lambda \right|$ DMD pixels. For DMD chips of approximately 1000 pixels this phase gradient is significant for $\left|\Delta \lambda / \lambda \right| > 0.1\%$. However, apart from this phase gradient the obtained field has a high fidelity until $\left| \Delta \lambda / \lambda \right| \approx 10\%$, at which point the field in the Fourier plane is so much displaced that the light starts to miss the spatial filter. 

A lookup table is used to make the connection between the desired target field at a superpixel and the combination of pixels within the superpixel that should be turned on in order to create that field. By using a lookup table the calculations needed to determine which DMD pixels to turn on are minimized and therefore the performance is optimized. We define a sufficiently fine square grid of possible target fields in the complex plane. We create a lookup table which contains for every point on this grid the nearest field the superpixel method can create as well as the combination of pixels that should be turned on in order to create this field. The size of the lookup table is chosen to be $855 \times 855$ points, about 100 times more dense than the set of possible target fields at superpixel size $n=4$. In our implementation it takes under 4 MB of memory to store the table. Loading the table and using it to look up a DMD pattern is done within a fraction of a second.

\section{Test field 1: LG$_{10}$ mode}

\begin{figure}[t]
\center{\includegraphics[width=14.5cm]{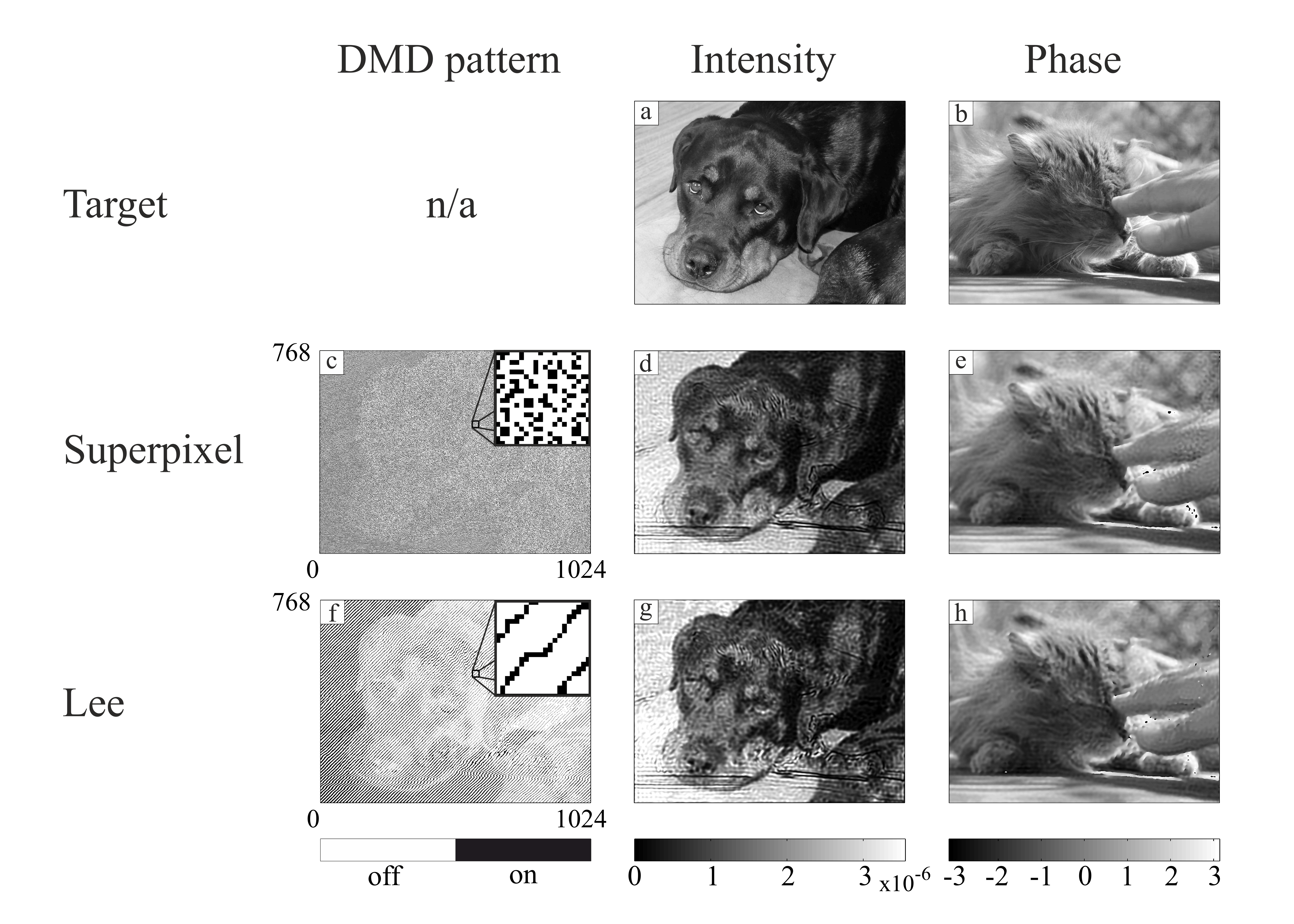}}
\caption{(a,b) Intensity and phase of a high resolution target field. (c) DMD pattern according to the superpixel method. Inset: zoom-in on $20\times 20$ DMD pixels. (d,e) Calculated intensity and phase using the superpixel method; $\delta_{\mathrm{superpixel}}^{\Delta\mathrm{k}}= 0.8\%$. (f) DMD pattern according to the Lee method using $k_{x}=k_{y}=\frac{2 \pi}{12}$ pixel$^{-1}$. Inset: zoom-in on $20\times 20$ DMD pixels. (g,h) Calculated intensity and phase using the Lee method; $\delta_{\mathrm{Lee}}^{\Delta\mathrm{k}}= 1.6\%$. Intensities are normalized to total intensity.}
\label{fig4}
\end{figure}

In order to test our method two test fields are constructed using superpixels of size $n=4$. The first test field is a LG$_{10}$ `donut' mode with an orbital angular momentum of $l=1$, where $l$ is the azimuthal mode number. These modes have many applications \cite{Padgett_advoptphoton_review}, including micromanipulation \cite{Puppe_prl,Allen_physreva}, imaging \cite{Hell_optlett_sted} and communication \cite{Gibson_optexpress}. The intensity and phase profiles of such a mode are shown in Figs. \ref{fig3}(a) and \ref{fig3}(b). In order to apply our superpixel method, we normalize the amplitude of the LG$_{10}$ mode to the maximum amplitude our method can create. For each superpixel we determine the pixel values using the lookup table. The resulting pattern on the DMD is shown in Fig. \ref{fig3}(c). For this low-resolution target field we tune the size $r$ of the spatial filter such that $\Delta k' = \frac{\pi}{100} \ \ \textrm{rad}\cdot \textrm{pixel}^{-1}$. This corresponds to a feature size of approximately $100\times 100$ DMD pixels and for our system this means $r=0.07$ mm. From the DMD pattern we calculate the resulting target field by first applying a fast Fourier transform for the first lens, then a multiplication with a circular mask for the spatial filter and finally a second fast Fourier transform for the second lens. The intensity and phase profiles of the obtained field are shown in Figs. \ref{fig3}(d) and \ref{fig3}(e). We observe an excellent match, apart from the phase in the corners which is not well defined as the intensity of the ideal LG$_{10}$ mode is negligible there. The fidelity of the superpixel method for this target field is calculated to be $F_{\mathrm{superpixel}}= 0.99993$. In other words, a fraction of only $\delta_{\mathrm{superpixel}} = 7\cdot 10^{-5}$ of the light goes to other modes.

The present reference method is Lee holography \cite{Lee1974}. Lee holography has two parameters: the size of the spatial filter and the spatial carrier frequency $\mathbf{k}$. The size of the spatial filter and therefore the system resolution are kept the same as when using the superpixel method. $\mathbf{k}$ is optimized to obtain maximum fidelity. The best result, which is obtained using $k_{x}=k_{y}=\frac{2 \pi}{30}$ pixel$^{-1}$, is shown in Figs. \ref{fig3}(f)--\ref{fig3}(h). Using this method the error $\delta_{\mathrm{Lee}} = 9\cdot 10^{-5}$. Both methods allow generation of a LG$_{10}$ mode with very high fidelity using a DMD of standard size. The superpixel method is most accurate, offering a $18\%$ reduction of error compared to Lee holography.

\section{Test field 2: Image quality}

Next, we consider a high resolution target field with uncorrelated intensity and phase. We choose a field which contains the picture of a dog in the intensity and a picture of a cat in the phase of the field, as shown in Figs. \ref{fig4}(a) and \ref{fig4}(b). Holographic methods are often used to project images and fully independent control over phase and amplitude of light is desired in many applications such as phase contrast microscopy \cite{mertz_2012}. Moreover, using this test field we show that the superpixel method can obtain a high resolution. We use superpixels of size $n=4$ and in order to allow for a high resolution we use an aperture size $r=0.9$ mm, which means the minimum feature size is $2\times 2$ superpixels. Any imaging system, and therefore any modulation method, has a finite resolution due to apertures in the system and the finite extent of the optics. This finite resolution leads to inevitable correlations between amplitude and phase of light fields. In particular, it is impossible to make very large phase gradients without the amplitude becoming zero. For the current test field and resolution, the theoretical maximum fidelity that can be achieved is given by $F_{\mathrm{theoretical}}^{\Delta\mathrm{k}}= \left| E_{\mathrm{target}}^* E_{\mathrm{target}}^{\Delta\mathrm{k}}\right|^2 = 0.955$, where $ E_{\mathrm{target}}^{\Delta\mathrm{k}}$ is the spatial bandwidth limited target field. 

\begin{figure}[t]
\center{\includegraphics[width=12cm]{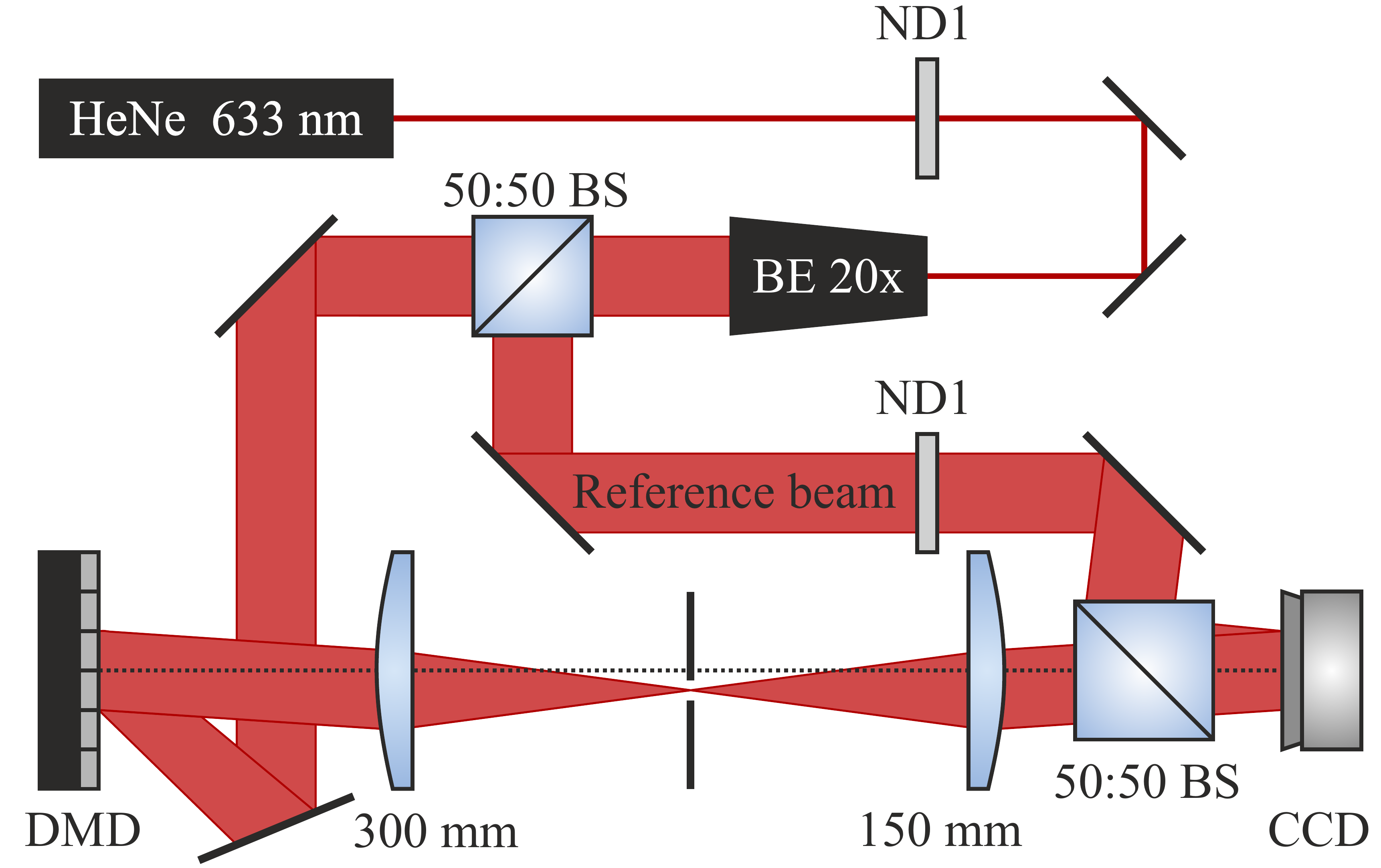}}
\caption{Experimental setup. DMD: ViALUX V4100, XGA resolution; CCD: AVT Dolphin F145-B; lenses: 2 inch achromats.}
\label{fig5}
\end{figure}

The DMD pattern and corresponding intensity and phase profiles that are obtained when using the superpixel method are shown in Figs. \ref{fig4}(c)--\ref{fig4}(e). We optimize Lee holography and find the optimum for $k_{x}=k_{y}=\frac{2 \pi}{12}$ pixel$^{-1}$. The resulting DMD pattern and obtained intensity and phase patterns are shown in Figs. \ref{fig4}(f)--\ref{fig4}(h). In both cases we observe some undesired ripples in the obtained intensity profile, because the steepest phase gradients in the target field cannot be resolved by the 8 DMD pixel resolution of the superpixel and Lee methods. We observe that the reconstructed intensity is more accurate when using the superpixel method. For the superpixel method we find a fidelity of $F_{\mathrm{superpixel}}= 0.947 = 0.992 F_{\mathrm{theoretical}}^{\Delta\mathrm{k}}$, showing that the fidelity is almost the theoretical maximum for the 8 pixel resolution. The error with respect to the bandwidth limited target is $\delta_{\mathrm{superpixel}}^{\Delta\mathrm{k}}= 0.8\%$. For Lee holography we find $\delta_{\mathrm{Lee}}^{\Delta\mathrm{k}}= 1.6\%$. The superpixel method offers a large improvement, reducing the error by $50\%$ compared to Lee holography.

\begin{figure}[t]
\center{\includegraphics[width=13.5cm]{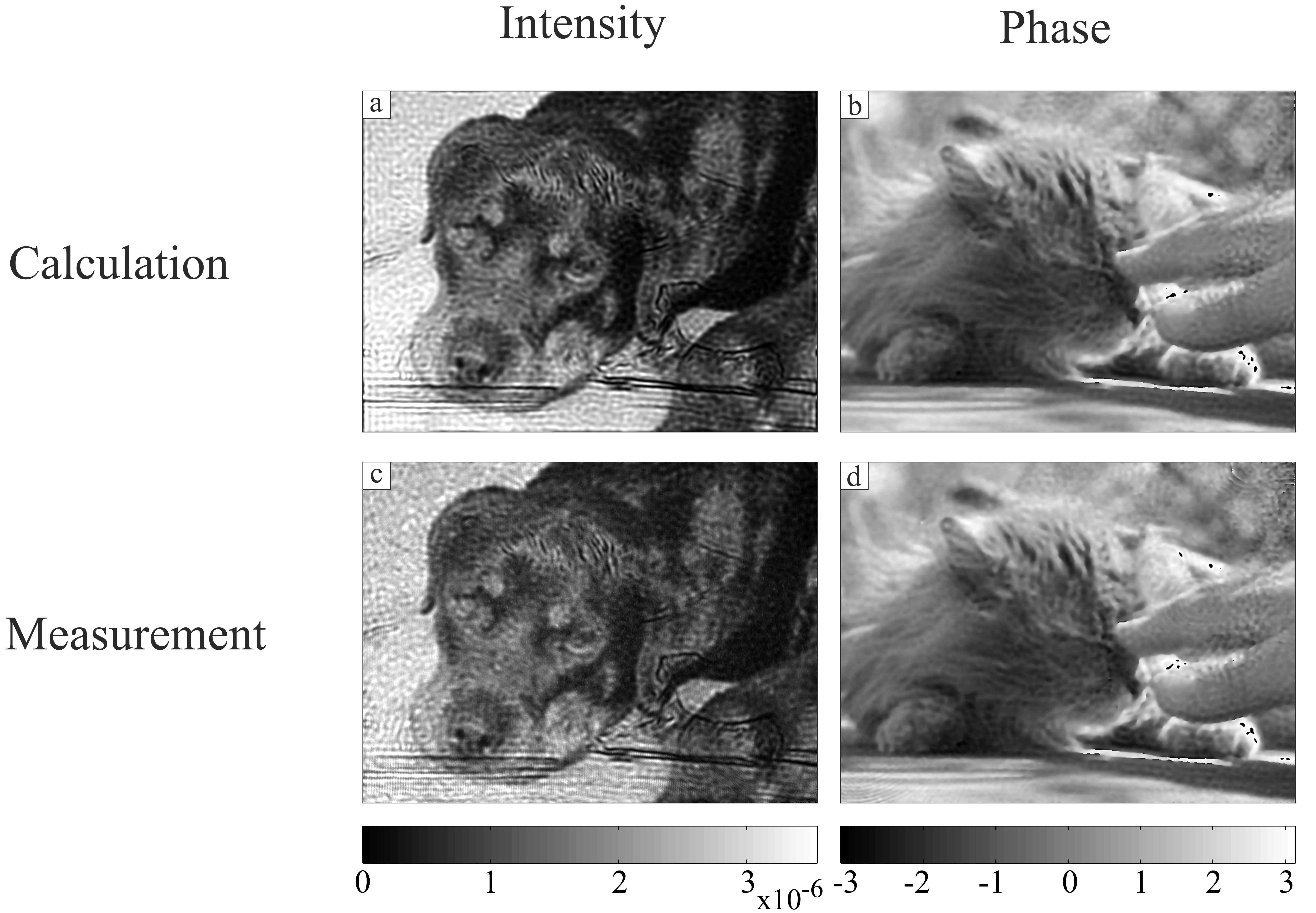}}
\caption{(a,b) Calculated intensity and phase using the superpixel method; $F_{\mathrm{superpixel}}= 0.99 F_{\mathrm{theoretical}}^{\Delta\mathrm{k}}$. (c,d) Measured intensity and phase using the superpixel method; $F_{\mathrm{superpixel, measured}}= 0.98 F_{\mathrm{theoretical}}^{\Delta\mathrm{k}}$. Intensities are normalized to total intensity.}
\label{fig6}
\end{figure}

We experimentally verified the fidelity of our superpixel method using the experimental setup shown in Fig. \ref{fig5} in combination with our publicly availabe control program implementing the superpixel method \cite{superpixel_software}. The constructed field is measured in the target plane on an AVT Dolphin F-145B CCD camera using off-axis digital holography \cite{Takeda1982}. The measured intensity is divided by the illumination intensity and from the measured phase we subtract the reference phase which is measured by constructing a plane wave. The measured field is shown in Fig. \ref{fig6}, along with the calculated field for comparison. We see that the measured field is almost identical to the calculated field. The fidelity of the measured field is $F_{\mathrm{superpixel, measured}}= 0.94 = 0.98 F_{\mathrm{theoretical}}^{\Delta\mathrm{k}}$, providing experimental proof that the superpixel method accurately constructs complex high resolution light fields. The small difference between the measured and calculated fidelity seems to be due to air flow causing a small phase error. 

\section{Origin of residual errors}
We identify two factors limiting the fidelity of the superpixel method. The first is the discrete approximation of the continuous target phases and amplitudes. In Fig. \ref{fig2} we observe that for the case of superpixels of size $4 \times 4$ each superpixel can create a large variety of complex fields. However, as in any modulation method, the modulation is discrete and there is a discretisation error. For test field 2 we compare the target field at each superpixel to the field that would ideally be obtained according to Fig. \ref{fig2}. We define $F_{\mathrm{discretisation}} =\left| E_{\mathrm{target}}^* E_{\mathrm{ideal superpixels}}\right|^2$, where $E_{\mathrm{ideal superpixels}}$ are the fields that can ideally be created at the superpixels and are taken directly from Fig. \ref{fig2}(b). We find $F_{\mathrm{discretisation}}=0.9995$, which means the discretisation error is an order of magnitude smaller than the total error. 

The remaining error can be explained by the displacement of pixels with respect to the center of the superpixel. In the calculation of the fields that can be constructed by a superpixel, as displayed in Fig. \ref{fig2}, pixel responses are assumed to be Airy disks located exactly at the center of the superpixel. Our assumption that the field constructed by a superpixel is an Airy disk positioned at the center of the superpixel only holds in that approximation. The effect of pixel displacement on fidelity depends on spatial correlations in the target field. For a field that is spatially uncorrelated at the scale of the system resolution, such as a speckle field with a speckle grain size of $2\times 2$ superpixels, the superpixel method achieves a fidelity $F=0.97$, as compared to $F=0.99$ for the more correlated test field 2. Resolution can be traded for fidelity: if the size of the spatial filter is reduced the resolution is reduced in exchange for a smaller relative pixel displacement and higher fidelity. For test field 1 we reach $F>0.9999$ at a resolution of $100\times 100$ DMD pixels.

Further reduction of the residual error may be possible by changing the position of the aperture in order to lift remaining degeneracy in the constructed fields, by taking into account the displacement error of the pixels when constructing the lookup table or by finding the optimal DMD setting iteratively to compensate for displacement errors. The simplicity of the current method is, however, a great advantage and may enable implementation in hardware such as Field Programmable Gate Arrays (FPGA).

\section{Conclusions}
We have demonstrated a superpixel based method to independently and accurately modulate the intensity and phase of light. Our method only requires very basic optics consisting of two lenses and a circular aperture, is very easy to align and highly robust to misalignment. The calculated modulation fidelity of our superpixel method exceeds 0.9999 for an LG$_{10}$ mode, using $768\times 768$ DMD pixels. Fidelity can be traded for resolution. We calculated and measured that at a resolution of $8\times 8$ DMD pixels per diffraction limited spot the modulation fidelity is in the order of 0.99 for our test image with uncorrelated intensity and phase. The superpixel method offers a modulation fidelity exceeding that of current methods and is expected to benefit the areas of imaging, holography, optical communication and optical micromanipulation.

\section*{Acknowledgments}
We thank Duygu Akbulut, Hasan Y\i lmaz, Henri Thyrrestrup, Michael J. Van De Graaff, Pepijn W.H. Pinkse, Ad Lagendijk and Willem L. Vos for discussions. This work is part of the research program of the Stichting voor Fundamenteel Onderzoek der Materie (FOM). A.P.M. acknowledges European Research Council grant no. 279248.

\end{document}